\documentclass{ieee}
\usepackage{graphicx}
\usepackage{amsmath}
\usepackage{array}
\usepackage{epsfig}

\begin{document}

\title{The Analysis of Large Order Bessel Functions in Gravitational
Wave Signals from Pulsars}

\author{F. A. Chishtie$^{1,3}$, S. R. Valluri$^{1,2,4}$, K. M. Rao$^{1}$, D. Sikorski$^{2}$ and T.
Williams$^{2}$\\
$^{1}$Department of Applied Mathematics,\\
University of Western Ontario, London, ON N6A 5B7\\
$^{2}$Department of Physics and Astronomy,\\
University of Western Ontario, London, ON N6A 3K7\\
Email: $^{3}$fchishti@uwo.ca\\
Email: $^{4}$valluri@uwo.ca}

\maketitle

\begin{abstract}
In this work, we present the analytic treatment of the large order
Bessel functions that arise in the Fourier Transform (FT) of the
Gravitational Wave (GW) signal from a pulsar. We outline several
strategies which employ asymptotic expansions in evaluation of
such Bessel functions which also happen to have large argument.
Large order Bessel functions also arise in the Peters-Mathews
model of binary inspiralling stars emitting GW and several
problems in potential scattering theory. Other applications also
arise in a variety of problems in Applied Mathematics as well as
in the Natural Sciences and present a challenge for High
Performance Computing(HPC).
\end{abstract}

\section{Introduction}
The detection of gravitational waves (GW) from astrophysical
sources is one of the most outstanding problems in experimental
gravitation today. Large laser interferometric gravitational wave
detectors like the LIGO, VIRGO, LISA, TAMA 300, GEO 600 and AIGO
are potentially opening a new window for the study of a vast and
rich variety of nonlinear curvature phenomena.

In recent works \cite{JVD96} we have analyzed the Fourier
transform (FT) of the Doppler shifted GW signal from a pulsar with
the use of the Plane Wave Expansion in Spherical Harmonics
(PWESH). Spherical-harmonic multipole expansions are used
throughout theoretical physics. The expansion of a plane wave in
spherical harmonics has a variety of applications not only in
quantum mechanics and electromagnetic theory \cite{MWIEEE}, but
also in many other areas. A number of researchers have used
spherical-harmonic expansions for a variety of problems in general
relativity, including problems where nonlinearity shows
up\cite{KThorne80}. The basis states in the PWESH expansion form a
complete set and facilitate such a study. It also turns out that
the consequent analysis of the Fourier Transform (FT) of the GW
signal from a pulsar has a very interesting and convenient
development in terms of the resulting spherical Bessel,
generalized hypergeometric function, the Gamma functions and the
Legendre functions. Both rotational and orbital motions of the
Earth and spindown of the pulsar can be considered in this
analysis which happens to have a nice analytic representation for
the GW signal in terms of the above special functions. The signal
can then be studied as a function of a variety of different
parameters associated with both the GW pulsar signal as well as
the orbital and rotational parameters. The numerical analysis of
this analytical expression for the signal offers a challenge for
fast and high performance parallel computation. The plane wave
expansion approach was also used by Bruce Allen and Adrian C.
Ottewill \cite{AO96} in their study of the correlation of GW
signals from ground-based GW detectors. They use the correlation
to search for anisotropies from stochastic background in terms of
the $l, m$ multipole moments. Our PWESH formalism enables a
similar study. Recent studies of the Cosmic Microwave Background
Explorer have raised the interesting question of the study of very
large multipole moments with angular momentum $l$ and its
projection $m$ going up to very large values of $l\sim1000$. Such
problems warrant an intensive analytic study supplemented by
numerical and parallel computation.

Since our FT depends on the Bessel function, a computational issue
arises due to large values of the index or order $n$ of the
function. In the GW form of the pulsar, the Doppler shifted
orbiting motion gives rise to Bessel functions $J_{n}(\frac{2 \pi
f_0 A \sin \theta}{c})$, where $\frac{2 \pi f_0 A \sin\theta}{c}$
is large for non-negligible angle $\theta$ as is shown in the
following section. Even for $\sin{\theta}\sim\frac{1}{1000}$, the
argument is large necessitating the consideration of large values
of $n$. The motivation of this work, is to extend the analysis in
Watson \cite{Watson} for large index, argument and overlapping
situations. Meissel \cite{M1} has made derivations for large order
Bessel functions both when the argument is smaller than the order
and vice versa. The asymptotics of these large order Bessel
functions are tricky in the sense that one runs into so-called
``transition" regions where such expansions fail. These regions
are values of the function when the argument is close to the given
order. As an application, we will address the phenomenological
situation of GW signal analysis of large order $n$ (which does
arise with combinations of $l$ and $m$) and supplement the
related computations with the presently derived results in a
forthcoming paper.

Captures of stellar-mass compact objects (CO) by massive black
holes are important capture sources for the Laser Interferometer
Space Antenna (LISA), the space based GW detector due to be
launched in about a decade\cite{PM}. Higher Harmonics of the
orbital frequency of the COs arise in the post Newtonian (PN)
capture GW model forms and contribute considerably to the total
signal to noise (S/N) ratio of the waveform. The GW form can be
decomposed into gravitational multipole moments which are treated
in the Fourier analysis of Keplerian eccentric orbits. The
radiation depends strongly on the orbital eccentricity $e$, and
Bessel functions $J_{n}(ne)$ are a natural consequence of the
analysis.

The calculation of partial derivatives of the potential scattering
phase shifts which often contain Bessel and Legendre functions of
large order angular momentum $l$, with respect to angular momentum
arise in a variety of scattering problems in atomic, molecular and
nuclear physics. In particular, large values of $l$ can arise in
rainbow, glory and orbit scattering. The analysis in our paper should help provide suitable approximations for large order and/or argument for the Bessel functions that arise in such problems.

\section{Fourier Transform of the GW signal}
The FT for the GW Doppler shifted pulsar signal \cite{JVD96} is
given as follows:

\begin{eqnarray}
\widetilde{h}(f)=S_{n l m}(\omega _{0},\omega
_{orb},T_{rE},n,l,m,A,R,k,\alpha ,\theta ,\phi)= \nonumber
\end{eqnarray}
\\
\begin{eqnarray}
{\sum_{n=-\infty}^{\infty}\sum_{l=0}^{\infty}\sum_{m=-l}^{l}\psi_0
\psi_1 \psi_2 \psi_3 \psi_4}
\end{eqnarray}
where
\begin{equation}
\psi_0(n,l,m,\alpha,\theta,\phi)= 4\pi i^{l}Y_{l m}(\theta ,\phi
)N_{l m}P_{l }^{m}(\cos \alpha )
\end{equation}
\begin{eqnarray}
\psi_1(n,\theta, \phi, T_{rE}, f_0,
A)=T_{rE}\sqrt{\frac{\pi}{2}}e^{-i\frac{2\pi f_{0}A}{c}\sin \theta
\cos \phi } i^{n}e^{-in\phi }\nonumber
\end{eqnarray}
\begin{eqnarray}
\times J_{n}\left(\frac{2\pi f_{0}A\sin \theta }{c}\right)
\end{eqnarray}
\begin{equation}
\psi_2(l,\omega_{orb}, \omega_{r}, n, m, R)=\left\{\frac{1-e^{i\pi
(l -B_{orb})R}}{1-e^{i\pi (l -B_{orb})}} \right\}
\frac{e^{-iB_{orb}\frac{\pi}{2}}}{2^{2l}}
\end{equation}

\begin{eqnarray}
\psi_3(k,l,m,n,\omega_{orb}, \omega_r)=k^{l+\frac{1}{2}}\nonumber\\
\times \frac{\Gamma \left(l +1\right) }{\Gamma \left(l
+\frac{3}{2}\right)\Gamma \left(\frac{l
+B_{orb}+2}{2}\right)\Gamma \left(\frac{l -B_{orb}+2}{2}\right)}
\end{eqnarray}

\begin{eqnarray}
\psi_4(k,l,m,n,\omega_{orb}, \omega_r)=_1F_{3}(l +1;l+\frac{3}{2}, \nonumber \\
\frac{l+B_{orb}+2}{2},\frac{l-B_{orb}+2}{2};\frac{-k^{2}}{16})
\end{eqnarray}

The angle $\alpha$ is the co-latitude detector angle and angles
$\theta$, $\phi$ are associated with the pulsar source. Here
$\omega_0=2\pi f_0$, $\omega_{orb}=\frac{2\pi}{T_{orb}}$
($T_{orb}=365$ days, $T_{rE} = 1$ day),
$B_{orb}=2\left(\frac{\omega-\omega_0}{\omega_r}+\frac{m}{2}+\frac{n
\omega_{orb}}{\omega_{rot}}\right)$, $k=\frac{4\pi f_0 R_E
\sin(\alpha)}{c}$ ($R_E$ is the radius of Earth, $c$ is the
velocity of light) and $A=1.5 \times 10^{11}$ meters is the
sun-earth distance.

\section{Extensions of Meissel's and Steepest Descent Expansions}
The Bessel function, of the type, $J_{\nu}(x)$ obeys the following
differential equation \cite{Watson},
\begin{equation}
z^2\frac{d^2J_{\nu}(\nu z)}{d z^2}+z\frac{dJ_{\nu}(\nu z)}{d
z}+\nu^2(1-z^2)J_{\nu}(\nu z)=0
\end{equation}
where the argument $x$ is parameterized by $\nu z$. If a function
$u(z)$ is introduced such that
\begin{equation}
J_{\nu}(\nu z)=\frac{\nu^{\nu}}{\Gamma (\nu+1)} \exp
\left\{\int_{}^{z} u(z) dz \right \}
\end{equation}
where $u(z)$ is a series in descending powers of $\nu$,
\begin{eqnarray}
u(z)=\nu u_0 + u_1 + \frac{u_2}{\nu} + \frac{u_3}{\nu^2}+
\frac{u_4}{\nu^3}+\frac{u_5}{\nu^4}+\frac{u_6}{\nu^5}+\frac{u_7}{\nu^6}\nonumber\\
+\frac{u_8}{\nu^7}+\frac{u_9}{\nu^8}+...
\end{eqnarray}
Substitution of this series and equation (8) in the differential
equation (7) yields the following expressions for $u_i(z)$,
$i=0...5$,
\begin{eqnarray}
u_0=\frac{\sqrt{1-z^2}}{z}, u_1=\frac{z}{2(1-z^2)},
u_2=-\frac{4z+z^2}{8(1-z^2)^{5/2}} \nonumber\\
u_3=\frac{4z+10z^3+z^5}{8(1-z^2)^{4}},
u_4=-\frac{64z+560z^3+456z^5+25z^7}{128(1-z^2)^{11/2}} \nonumber\\
u_5=\frac{16z+368z^3+924z^5+347z^7+13z^9}{32(1-z^2)^{7}} \nonumber
\end{eqnarray}
Hence, by integrating $u_i$, and substituting in Equation (8) we
arrive at Meissel's \textit{First} expansion \cite{M1}, which is
valid for the case when the argument is less than the order $\nu$.
We do not list $u_6,u_7,u_8$ and $u_9$ as one can obtain these
straightforwardly from their respective integrals shown below.
These results are expressed as,

\begin{equation}
J_{\nu}(\nu z)=\frac{(\nu z)^{\nu} \exp (\nu
\sqrt{1-z^2})\exp(-V_{\nu})}{e^{\nu}\Gamma (\nu+1)
(1-z^2)^{1/4}[1+\sqrt{1-z^2}]^{\nu}}
\end{equation}

where,
\begin{equation}
V_{\nu}=V_1+V_2+V_3+V_4+V_5+V_6+V_7+V_8+...
\end{equation}
and,
\begin{eqnarray}
V_1=\frac{1}{24\nu}\left(\frac{2+3z^2}{(1-z^2)^{3/2}}-2\right),V_2=-\frac{4z^2+z^4}{16\nu^2(1-z^2)^3}\nonumber
\end{eqnarray}

\begin{eqnarray}
V_3=-\frac{1}{5760\nu^3}\left(\frac{16-1512z^2-3654z^4-375z^6}{(1-z^2)^{9/2}}-16\right)\nonumber
\end{eqnarray}

\begin{eqnarray}
V_4=-\frac{32z^2+288z^4+232z^6+13z^8}{128\nu^{4}(1-z^2)^6}\nonumber
\end{eqnarray}

\begin{eqnarray}
V_5=-\frac{1}{322560\nu^5(1-z^2)^{15/2}}(67599\,{z}^{10}+1914210\,{z}^{8}\nonumber\\
+4744640\,{z}^{6}+1891200\,{z}^{4}+78720\,{z}^{2}+256)+\frac{1}{1260\nu^5}\nonumber
\end{eqnarray}

\begin{eqnarray}
V_6=\frac{z^2}{192(1-{z}^{2})^{9}{\nu}^{6}}(48+2580{z}^{2}+14884{z}^{4}\nonumber\\
+17493{z}^{6}+4242{z}^{8}+103{z}^{10})\nonumber
\end{eqnarray}

\begin{eqnarray}
V_7=-\frac{(1-z^2)^{-21/2}}{3440640\nu^7}(881664{z}^{2}+99783936{z}^{4}\nonumber
\end{eqnarray}
\begin{eqnarray}
+1135145088{z}^{6}+2884531440{z}^{8}+1965889800{z}^{10}\nonumber\\
+318291750{z}^{12}+5635995{z}^{14}-2048)-\frac{1}{1680\nu^7}\nonumber
\end{eqnarray}

\begin{eqnarray}
V_8={\frac{z^2}{4096(1-{z}^{2})^{12}{\nu}^{8}}}(1024+248320{z}^{2}+5095936{z}^{4}\nonumber\\
+24059968{z}^{6}+34280896{z}^{8}+15252048{z}^{10}\nonumber\\
+1765936{z}^{12}+23797{z}^{14})\nonumber
\end{eqnarray}
Hence we have actually increased Meissel's analysis by two orders.
Using symbolic packages these orders were computed and higher
terms should pose no problem if the application requires higher
accuracy.

For the case when the argument is larger than the index, Meissel
used the parametrization $z=\sec{\beta}$ \cite{M1}, and we shall
term it as his \textit{Second} expansion. Hence,
\begin{equation}
J_{\nu}(\nu \sec{\beta})=\sqrt{\frac{2 \cot{\beta}}{\nu \pi}}
e^{-P_{\nu}}\cos\left(Q_{\nu}-\frac{1}{4}\pi \right)
\end{equation}

where $P_{\nu}$ is given as,

\begin{equation}
P_{\nu}=P_1+P_2+P_3+P_4+...
\end{equation}
where

\begin{eqnarray}
P_{1}=\frac{\cot^{6}\beta}{16\nu^2}\left(4\sec^2{\beta}+\sec^4{\beta}\right)\nonumber
\end{eqnarray}

\begin{eqnarray}
P_{2}=-\frac{\cot^{12}\beta}{128\nu^4}(32\sec^2{\beta}+288\sec^4{\beta}+232\sec^6{\beta}\nonumber\\
+13\sec^8{\beta})\nonumber
\end{eqnarray}

\begin{eqnarray}
P_{3}=\frac{\cot^{18}\beta}{192\nu^6}(48\sec^2{\beta}+2580\sec^4{\beta}+14884\sec^6{\beta}\nonumber\\
+17493\sec^8{\beta}+4242\sec^{10}{\beta}+103\sec^{12}{\beta})\nonumber
\end{eqnarray}

\begin{eqnarray}
P_{4}=\frac{\cot^{24}\beta\sec^2{\beta}}{4096\nu^8}(1024+248320\sec^2{\beta}+5095936\sec^4{\beta}\nonumber
\end{eqnarray}
\begin{eqnarray}
+24059968\sec^6{\beta}+34280896\sec^8{\beta}+15252048\sec^{10}{\beta}\nonumber\\
+1765936\sec^{12}{\beta}+23797\sec^{14}{\beta})\nonumber
\end{eqnarray}

and $Q_{\nu}$ is given as,

\begin{equation}
Q_{\nu}=Q_1+Q_2+Q_3+Q_4+...
\end{equation}
and,
\begin{eqnarray}
Q_1=\nu(\tan{\beta}-\beta)-\frac{\cot^{3}\beta}{24\nu}\left(2+3\sec^2{\beta}\right)\nonumber
\end{eqnarray}
\begin{eqnarray}
Q_2=-\frac{\cot^{9}\beta}{5760\nu^3}(16-1512\sec^2{\beta}-3654\sec^4{\beta}\nonumber\\
-375\sec^6{\beta})\nonumber
\end{eqnarray}
\begin{eqnarray}
Q_3=-\frac{\cot^{15}\beta}{322560\nu^5}(256+78720\sec^2{\beta}+1891200\sec^4{\beta}\nonumber\\
+4744640\sec^6{\beta}+1914210\sec^{8}{\beta}+67599\sec^{10}{\beta})\nonumber
\end{eqnarray}
\begin{eqnarray}
Q_4=-\frac{\cot^{21}\beta}{3440640\nu^7}(881664\sec^2{\beta}+99783936\sec^4{\beta}\nonumber
\end{eqnarray}
\begin{eqnarray}
+1135145088\sec^6{\beta}+2884531440\sec^8{\beta}+1965889800\sec^{10}{\beta}\nonumber\\
+318291750\sec^{12}{\beta}+5635995\sec^{14}{\beta}-2048)\nonumber
\end{eqnarray}
It should be remarked that we disagree with Meissel's result for
$P_3$ in the last four terms. However, we obtain perfect agreement
with the rest of his results \cite{M1}. We have improved on his
result by using $V_7$, $V_8$ to obtain $P_4$ and $Q_4$. Hence, we
have increased the accuracy of this expansion by at least one
order from Meissel's earlier result. Again, higher order results
are easily obtainable and are available if needed.

In Figures 1 and 2 we have plotted these expansions in the regions
they are expected to fail. These are the so called ``transition"
regions, where each expansion approaches a singularity (as the
order equals the argument). For the computationally motivated (we
can compute exact values of Bessel functions with ease) case of
the $\nu=300$, we note the following. Fig.1 indicates the onset of
breakdown in the \textit{First} expansion for argument values
around and larger than 290. Similarly, Figure 2, indicates a
similar breakdown starting around the values 300 and persisting
till 310. Hence, the values outside these regions of breakdown or
transition regions are well covered by Meissel's expansions.
However, the issue as to deal with these regions need to be
addressed via separate methods, which will be addressed in more
detail in Section IV. The CPU time for these approximations was
less than 0.01 seconds per value on a 2.4 GHz Pentium IV processor
running MAPLE version 9. The ``exact" MAPLE solver took somewhere
between 0.03 to 0.08 seconds to compute each value. Clearly, there
is a lot more computational speed in using a few terms present in
these expansions. As an application, it should be noted that
values of this order are applicable to the Peters-Mathews model of
gravitational radiation from binary inspiralling stars \cite{PM}.

\begin{figure}
\centering \epsfig{file=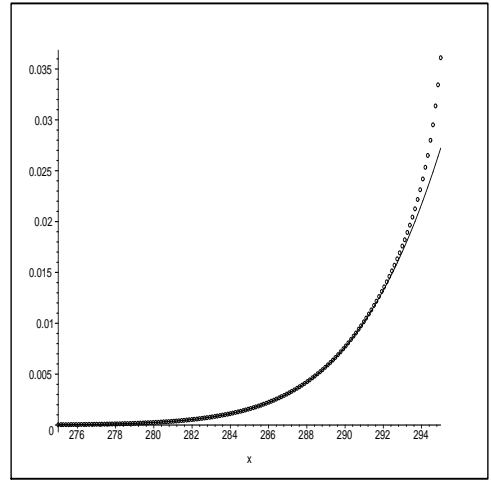,width= 2.5 in,
height= 2.5 in, angle=270} \caption{Meissel's \textit{First}
expansion and actual Bessel function graphed for argument $x$ and
order $\nu=300$ near the transition region. Solid line indicates
actual Bessel function values and circles indicate values given by
the expansion.} \label{fig_three}
\end{figure}

\begin{figure}
\centering \epsfig{file=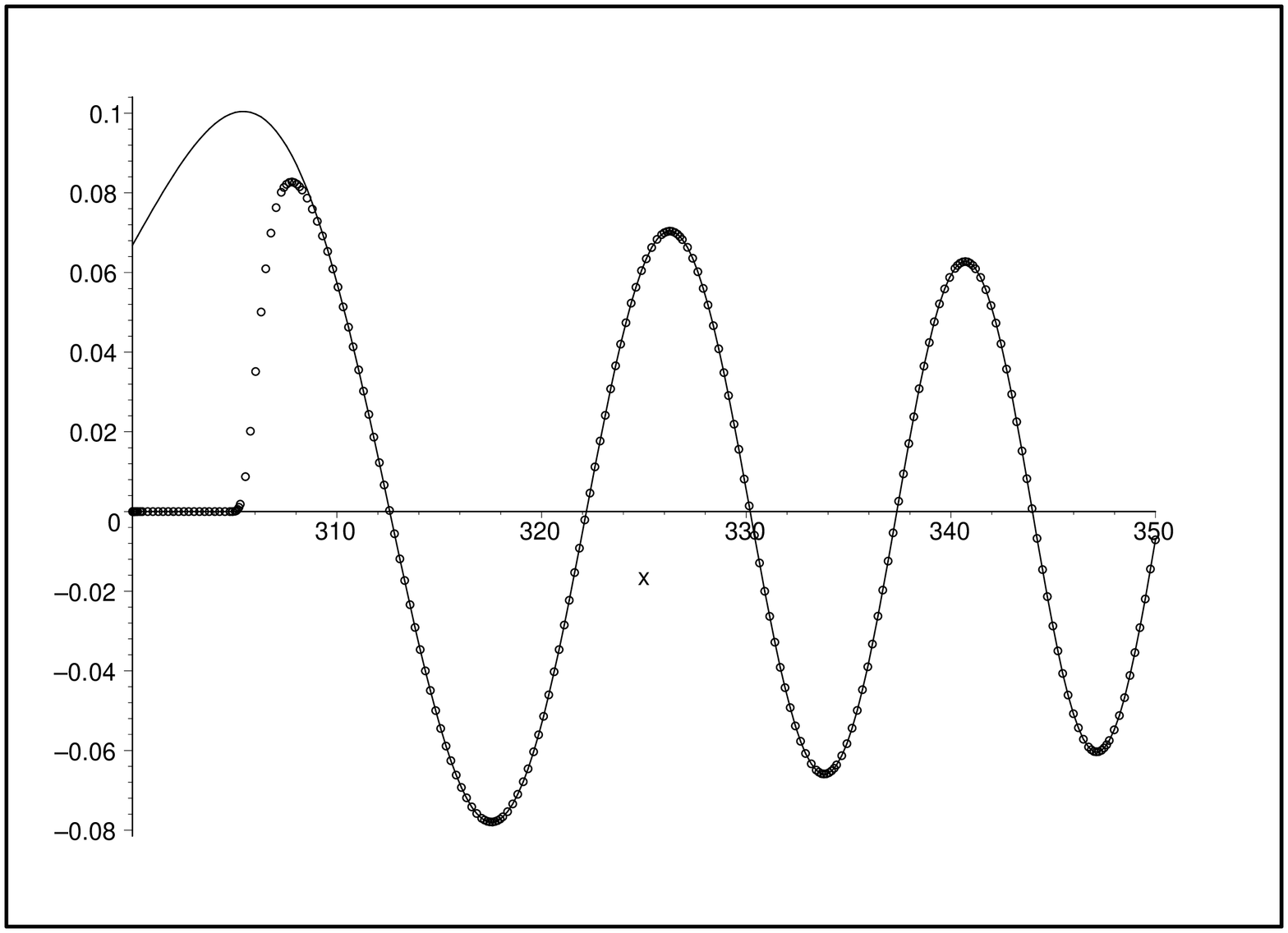,width= 2.5 in,
height= 2.5 in, angle=270} \caption{Meissel's \textit{Second}
expansion and actual Bessel function graphed for argument $x$ and
order $\nu=300$ near the transition region. Solid line indicates
actual Bessel function values and circles indicate values given by
the expansion.} \label{fig_two}
\end{figure}

For the case when the argument equals the index, we extend
Meissel's \textit{Third} expansion \cite{M1} by two orders as
follows:
\begin{equation}
J_{n}(n)\sim\frac{1}{\pi}\sum_{m=0}^{\infty} \lambda_{m}
\Gamma\left(\frac{2m}{3} + \frac{4}{3}
\right)\left(\frac{6}{n}\right)^{\frac{2}{3}m +
\frac{1}{3}}\cos\pi (\frac{m}{3} + \frac{1}{6})
\end{equation}
where the terms, $\lambda_m$ ($m=0,1,2,..7$), are given by,

\begin{eqnarray}
\lambda_0=1,\lambda_1=\frac{1}{60},\lambda_2=\frac{1}{1400},\lambda_3=\frac{1}{25200},\nonumber
\end{eqnarray}
\begin{eqnarray}
\lambda_4=\frac{43}{17248000},\lambda_5=\frac{1213}{7207200000},\lambda_6=\frac{681563}{5721073600000},\nonumber
\end{eqnarray}
\begin{eqnarray}
\lambda_7=\frac{63319}{726485760000000}
\end{eqnarray}
We observe that inclusion of the higher order terms leads to 10
decimal accuracy compared to actual values of large order Bessel
functions.

The method of steepest descents was employed by Debye in
\cite{DBY}. For the case when the argument is less than the order,
he obtained,
\begin{equation}
J_{\nu}(\nu sech(\alpha))\sim
\frac{e^{\nu(\tanh{\alpha}-\alpha)}}{\sqrt{2 \pi
\nu\tanh{\alpha}}}
\sum_{m=0}^{\infty}\frac{\Gamma(m+\frac{1}{2})}{\Gamma(\frac{1}{2})}\frac{A_m}{(\frac{1}{2}\nu
\tanh{\alpha})^{m}}
\end{equation}
where,
\begin{eqnarray}
A_0=1, A_1=\frac{1}{8}-\frac{5}{24}\coth^2{\alpha} \nonumber
\end{eqnarray}
\begin{eqnarray}
A_2=\frac{3}{128}-\frac{77}{576}\coth^2{\alpha}+\frac{385}{3456}\coth^4{\alpha}\nonumber
\end{eqnarray}
\begin{eqnarray}
A_3=\frac{5}{1024}-\frac{1521}{25600}\coth^2{\alpha}+\frac{17017}{138240}\coth^4{\alpha}\nonumber\\
-\frac{17017}{248832}\coth^6{\alpha}\nonumber
\end{eqnarray}
\begin{eqnarray}
A_4=\frac{11513}{92897280}-\frac{21023}{9953280}\coth^2{\alpha}+\frac{138919}{19906560}\coth^4{\alpha}\nonumber\\
-\frac{49049}{5971968}\coth^6{\alpha}+\frac{230945}{71663616}\coth^8{\alpha}\nonumber
\end{eqnarray}
Following this method, we have computed two higher orders $A_3$
and $A_4$, using symbolic computation.

For the case when the argument is larger than the order, Debye
obtains the following expansion:
\begin{eqnarray}
J_{\nu}(\nu\sec{\beta})\sim\sqrt{\frac{2}{\pi\nu\tan{\beta}}}[\cos\left(\nu\tan{\beta}-\nu\beta-\frac{1}{4}\beta\right)\nonumber
\end{eqnarray}
\begin{eqnarray}
\times\sum_{m=0}^{\infty}(-1)^{m}\frac{\Gamma(m+\frac{1}{2})}{\Gamma(\frac{1}{2})}\frac{A_{2m}}{(\frac{1}{2}\nu\tanh{\alpha})^{2m}}+\sin(\nu\tan{\beta}\nonumber
\end{eqnarray}
\begin{eqnarray}
-\nu\beta-\frac{1}{4}\beta)\sum_{m=0}^{\infty}(-1)^{m}\frac{\Gamma(2m+\frac{3}{2})}{\Gamma(\frac{1}{2})}\frac{A_{2m+1}}{(\frac{1}{2}\nu
\tanh{\alpha})^{2m+1}}]
\end{eqnarray}

where,
\begin{eqnarray}
A_0=1, A_1=\frac{1}{8}+\frac{5}{24}\cot^2{\beta}\nonumber
\end{eqnarray}
\begin{eqnarray}
A_2=\frac{3}{128}+\frac{77}{576}\cot^2{\beta}+\frac{385}{3456}\cot^4{\beta}\nonumber
\end{eqnarray}
\begin{eqnarray}
A_3=\frac{5}{1024}+\frac{1521}{25600}\cot^2{\beta}+\frac{17017}{138240}\cot^4{\beta}\nonumber\\
+\frac{17017}{248832}\cot^6{\beta}\nonumber
\end{eqnarray}
\begin{eqnarray}
A_4=\frac{11513}{92897280}+\frac{21023}{9953280}\cot^2{\beta}+\frac{138919}{19906560}\cot^4{\beta}\nonumber\\
+\frac{49049}{5971968}\cot^6{\beta}+\frac{230945}{71663616}\cot^8{\beta}\nonumber
\end{eqnarray}

Again, we have extended Debye's result by two higher orders by
obtaining $A_3$ and $A_4$. However, due to the nature of this
method we could not obtain reliable results that spanned in a
generally predictable direction. Accuracy was limited to the
region of the stationary phase as expected and hence, we recommend
Meissel's expansions to be more reliable (except of course in the
``transition" region) than the method of steepest descent.

\section{Transitional regions: Contour Integration and extension of $\epsilon$ expansion}
To address the issues related to computation for large order
Bessel functions in the transition regions we present two methods
that are geared to work in such domains. Firstly, we present the
results by Watson, \cite{Watson}. For the case of the argument
being less than the order, he obtained via use of contour
integration,
\begin{eqnarray}
J_{\nu}(\nu sech(\alpha))= \frac{\tanh{\alpha}}{\pi
\sqrt{3}}\exp\left[\nu\left(\tanh{\alpha}+\frac{1}{3}\tanh^3{\alpha}-\alpha\right)\right]\nonumber\\
\times
K_{\frac{1}{3}}\left(\frac{1}{3}\nu\tanh^3{\alpha}\right)\nonumber
\end{eqnarray}
\begin{eqnarray}
+3\theta_{1}\nu^{-1}\exp[\nu(\tanh{\alpha}-\alpha)]
\end{eqnarray}
where $\theta_1<1$. Similarly, for the case when the argument is
greater than the order, he derived the following:
\begin{eqnarray}
J_{\nu}(\nu
\sec{\beta})=\frac{1}{3}\tan{\beta}\cos\left[\nu\left(\tan{\beta}-\frac{1}{3}\tan^3{\beta}-\beta\right)\right]\times\nonumber
\end{eqnarray}
\begin{eqnarray}
\left(J_{-\frac{1}{3}}+J_{\frac{1}{3}}\right)+3^{-\frac{1}{2}}\tan{\beta}\sin\left[\nu\left(\tan{\beta}-\frac{1}{3}\tan^3{\beta}-\beta\right)\right]\times\nonumber\\
\left(J_{-\frac{1}{3}}-J_{\frac{1}{3}}\right)\nonumber
\end{eqnarray}
\begin{eqnarray}
+24\theta_{2}\nu^{-1}
\end{eqnarray}

where $\theta_2<1$ and the argument for the Bessel functions
$J_{\pm\frac{1}{3}}$ is $\frac{1}{3}\tan^{3}{\beta}$. The great
advantage of these formulae is that they have error bounds given.
However, these extensions are not trivial as this involves solving
extensions to Airy-type integrals, for which we do not presently
have closed form answers. The other issue with these formulae is
that they are themselves given in fractional Bessel function form
which would pose computational problems once the arguments
involved are large.

On the other hand, Debye \cite{DBY}, introduced, what we will term
as ``$\epsilon$ expansion". The idea is motivated by introducing a
small parameter $\epsilon$, such that $\nu=z(1-\epsilon)$, where
$\nu$ denotes the order and $z$ is the argument of the Bessel
function.
\begin{equation}
J_{\nu}(z)\sim\frac{1}{3\pi}\sum_{m=0}^{\infty} B_{m}(\epsilon z)
\sin\frac{1}{3}(m+1)\pi\cdot
\frac{\Gamma(\frac{1}{3}m+\frac{1}{3})}{(\frac{1}{6}z)^{\frac{1}{3}(m+1)}}
\end{equation}

We have extended this analysis by 5 orders and the terms
$B_m(\epsilon z), m=0,1,2,..15$, are given as,

\begin{eqnarray}
B_0(\epsilon z)=1, B_1(\epsilon z)=\epsilon z, B_3(\epsilon
z)=\frac{1}{6}\epsilon^3 z^3 -\frac{1}{15} \epsilon z \nonumber
\end{eqnarray}

\begin{eqnarray}
B_4(\epsilon z)=\frac{1}{24}\epsilon^4 z^4 -\frac{1}{24}
\epsilon^2 z^2 + \frac{1}{280}\nonumber
\end{eqnarray}

\begin{eqnarray}
B_6(\epsilon z)={\frac {1}{720}}\,{z}^{6}{\epsilon}^{6}-{\frac
{7}{1440}}\,{z}^{4}{ \epsilon}^{4}+{\frac
{1}{288}}\,{z}^{2}{\epsilon}^{2}-{\frac {1}{3600} }
\end{eqnarray}

\begin{eqnarray}
B_7(\epsilon z)={\frac {1}{5040}}\,{z}^{7}{\epsilon}^{7}-{\frac
{1}{900}}\,{z}^{5}{ \epsilon}^{5}+{\frac
{19}{12600}}\,{z}^{3}{\epsilon}^{3}-{\frac {13}{
31500}}\,z\epsilon \nonumber
\end{eqnarray}

\begin{eqnarray}
B_{9}(\epsilon z)={\frac
{1}{362880}}\,{z}^{9}{\epsilon}^{9}-{\frac {1}{30240}}\,{z}^{7}
{\epsilon}^{7}+{\frac
{71}{604800}}\,{z}^{5}{\epsilon}^{5}\nonumber\\
-{\frac{121}{907200}}\,{z}^{3}{\epsilon}^{3}+{\frac{7939}{232848000}}\,z\epsilon\nonumber
\end{eqnarray}

\begin{eqnarray}
B_{10}(\epsilon z)={\frac
{1}{3628800}}\,{z}^{10}{\epsilon}^{10}-{\frac {11}{2419200}}\,{
z}^{8}{\epsilon}^{8}+{\frac
{143}{6048000}}\,{z}^{6}{\epsilon}^{6}\nonumber\\
-{ \frac{803}{18144000}}\,{z}^{4}{\epsilon}^{4}+{\frac
{43}{1728000}}\,{ z}^{2}{\epsilon}^{2}-{\frac
{1213}{655200000}}\nonumber
\end{eqnarray}

\begin{eqnarray}
B_{12}(\epsilon z)={\frac
{1}{479001600}}\,{z}^{12}{\epsilon}^{12}-{\frac{13}{217728000}}\,{z}^{10}{\epsilon}^{10}+\nonumber\\
{\frac
{299}{508032000}}\,{z}^{8}{\epsilon}^{8}-{\frac{377}{155520000}}\,{z}^{6}{\epsilon}^{6}+{\frac{337207}{83825280000}}\,{z}^{4}{\epsilon}^{4}\nonumber\\
-{\frac{59503}{27941760000}}\,{z}^{2}{\epsilon}^{2}+{\frac{151439}{977961600000}}\nonumber
\end{eqnarray}

\begin{eqnarray}
B_{13}(\epsilon z)={\frac
{1}{6227020800}}\,{z}^{13}{\epsilon}^{13}-{\frac{1}{171072000}}\,{z}^{11}{\epsilon}^{11}+\nonumber\\
{\frac{11}{145152000}}\,{z}^{9}{\epsilon}^{9}-{\frac{47}{108864000}}\,{z}^{7}{\epsilon}^{7}+{\frac{25853}{23950080000}}\,{z}^{5}{\epsilon}^{5}\nonumber\\
-{\frac{266303}{259459200000}}\,{z}^{3}{\epsilon}^{3}+\frac{169039}{698544000000}\,z\epsilon\nonumber
\end{eqnarray}

\begin{eqnarray}
B_{15}(\epsilon z)={\frac
{1}{1307674368000}}\,{z}^{15}{\epsilon}^{15}-{\frac{1}{23351328000}}\,{z}^{13}{\epsilon}^{13}\nonumber\\
+{\frac{113}{125737920000}}\,{z}^{11}{\epsilon}^{11}-{\frac{17}{1905120000}}\,{z}^{9}{\epsilon}^{9}\nonumber\\
+{\frac{76841}{1760330880000}}\,{z}^{7}{\epsilon}^{7}-{\frac{37021}{371498400000}}\,{z}^{5}{\epsilon}^{5}\nonumber\\
+{\frac{5141933}{57210753600000}}\,{z}^{3}{\epsilon}^{3}-{\frac{16720141}{810485676000000}}\,z\epsilon\nonumber
\end{eqnarray}

Terms $B_{3m-1}$, $m=1,2...$  do not contribute in eqn. (21) due
to the periodicity of the sine function. With symbolic
computation, we are able to generate higher orders if needed.

To illustrate the applicability and issues of both these methods
to the transition region, we present Figures 3 and 4, which are
plotted for the problematic regions (when the order is $\nu=300$)
in Figures 1 and 2. Both methods show remarkable ability in
capturing the functions in the domains of interest. In Figure 3,
the $\epsilon$ expansion starts working at values at 286 and
Watson's formula works to even a larger domain. Similarly, in
Figure 4, both the methods indicate success in regions where
Meissel's expansions fail. This starts at values of the argument,
and works up to $x=316$ for the $\epsilon$ expansion whereas,
again, the domain of Watson's formula is much greater. The reasons
for lesser range of the $\epsilon$ expansion can be attributed to
the fact that it is a power series compared to Watson's formula
which actually depends on fractional Bessel functions themselves.
Further, the $\epsilon$ expansion depends crucially on the size of
the parameter, which is connected with the order one is working
with. However, the reason why we will persist with this method is
that it will be more applicable when the argument of the Bessel
function is quite large.
\begin{figure}
\centering \epsfig{file=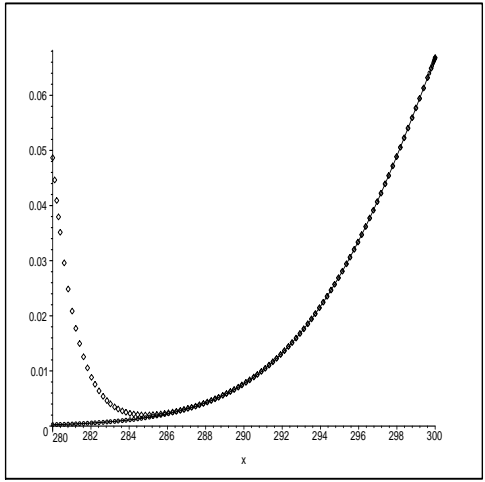,width= 2.5 in, height=
2.5 in, angle=270} \caption{Comparison of $\epsilon$ expansion and
Watson's formulae for argument $x<300$ and order $\nu=300$.Solid
line indicates exact Bessel function values, diamonds represent
$\epsilon$ expansion and circles indicate values given by Watson's
formula.} \label{fig_three}
\end{figure}

\begin{figure}
\centering \epsfig{file=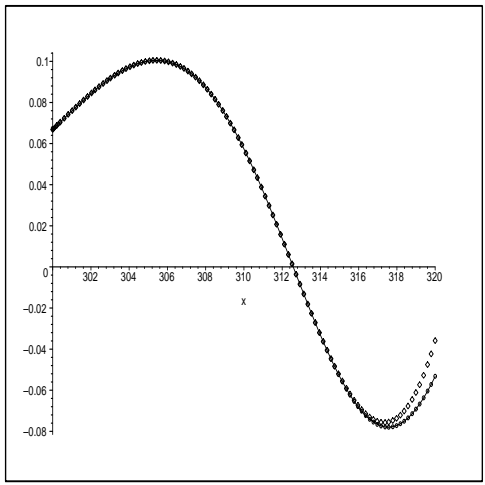,width= 2.5 in, height=
2.5 in, angle=270} \caption{Comparison of $\epsilon$ expansion and
Watson's formula in the transition region for argument $x>300$ and
order $\nu=300$. Solid line indicates exact Bessel function
values, diamonds represent $\epsilon$ expansion and circles
indicate values given by Watson's formula.} \label{fig_four}
\end{figure}

To illustrate the type of values a GW pulsar FT would require, we
present Figures 5 and 6. Here, we choose a very large order (yet
realistic phenomenologically) for the Bessel function, which is 1
million. Also, in such a scenario, we would be looking at values
greater than one million, hence Meissel's second expansion along
with the appropriate Watson's formula (eq. 20) will be put to use.
We were not able to make exact comparison, obviously due to
massive computer times required. In this regard, the problem of
``exact" Bessel functions presents a genuine challenge to SHARCNET
(Shared Hierarchical Academic Research Cluster Network) and HPC in
general. In Figure 5, we observe strong evidence that the proposed
asymptotic expansions are appropriate for GW signal analysis.
Here, we note the transition region starting at values of the
argument at 1,000,000 and going up to 1,000,200. In this region,
both the $\epsilon$ expansion and Watson's formula almost coincide
with each other. As usual, the $\epsilon$ expansion breaks down
earlier, however, all three methods coincide in a certain region
indicating that we have consistent methods that work for values
relevant to GW analysis. Meissel's expansion is fairly easy to
implement computationally and indicates good stability for rather
large values of the argument. This is illustrated in Figure 6,
where we plot this expansion for values ranging from 1,000,200 to
32,500,000, which are relevant for GW phenomenology. This appears
as a black band and is a continuous function which indicates
oscillations tightly bunched together. It is noteworthy that the
method is stable and shows consistent behaviour over an extreme
range of values for the argument. The CPU time consumed by each of
the points, on the average took less than 0.01 seconds on MAPLE.
The Bessel utility in MAPLE crashed repeatedly after 15-30 minutes
on the same system described above. It should be remarked that
Watson's formula lacks in this capacity as it depends on
fractional Bessel functions itself, which will provide
computational challenge for such values. A detailed analysis
regarding computational advantage over exact computation will be
addressed in a later work. It is aimed to not only address the
question of GW analysis but will deal with general computational
issues regarding large order Bessel functions.
\begin{figure}
\centering \epsfig{file=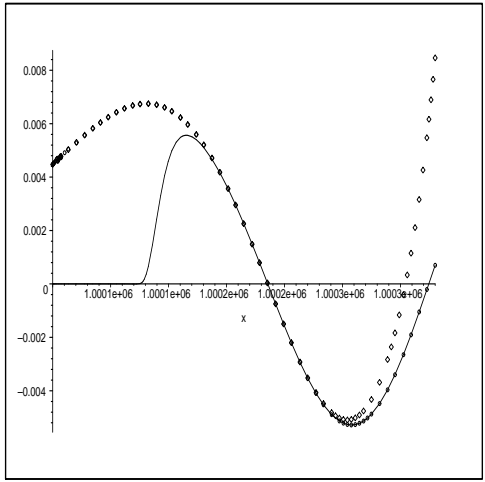,width= 2.5 in,
height= 2.5 in, angle=270} \caption{Comparison of Meissel's
\textit{Second} expansion, $\epsilon$ expansion and Watson's
formulae for argument $x>1,000,000$ and order $\nu=1,000,000$.
Solid line indicates Meissel's \textit{Second} expansion values,
diamonds represent $\epsilon$ expansion and circles indicate
values given by Watson's formula.} \label{fig_five}
\end{figure}

\begin{figure}
\centering \epsfig{file=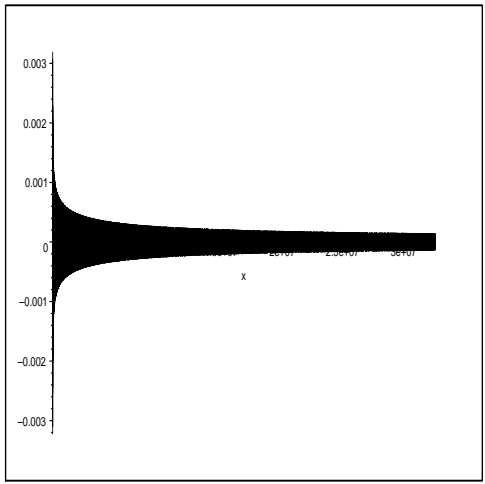,width= 2.5
in, height= 2.5 in, angle=270} \caption{Plot of Meissel's
\textit{Second} expansion for argument, $x$ ranging from 1,000,200
to 32,500,000 for order 1,000,000.} \label{fig_six}
\end{figure}

\section{Conclusion}
In this present work, we have given an extended asymptotic
analysis for the large order and argument Bessel functions. This
analytically improves the earlier pioneering works of Meissel,
Airey, Debye and Watson. These extensions should be of possible
use not only in GW signal analysis, but also in a variety of
problems in Engineering and the Sciences where the ubiquitous
Bessel functions are encountered.

\section*{Acknowledgments}
We are deeply grateful to SHARCNET for valuable grant support that
made this study feasible. We are also indebted to Drs. Nico Temme
(CWI, Amsterdam), Walter Gautschi (Purdue U.), D.G.C. McKeon (U.
Western Ontario), Tom Prince (JPL, Pasadena) and the referee for
valuable suggestions.


\begin{thebibliography}{99}

\bibitem{JVD96}Kanti Jotania, S.R. Valluri and Sanjeev V. Dhurandhar, ``A Study of the Gravitational
Wave Form from Pulsars", \textit{Astron. Astrophys.} vol. 306, pp.
317, 1996; S.R. Valluri, F.A. Chishtie, J.J. Drozd, R.G. Biggs,
M.Davison, S. V. Dhurandhar and B.S. Sathyaprakash, ``A study of
the Gravitational Wave Form from Pulsars", \textit{Class. Quant.
Grav.} vol.19 pp. 1327-1334, 2002, Erratum-ibid. vol. 19, pp.
4227-4228, 2002.

\bibitem{MWIEEE}MacPhie, R. H. and Wu, K.-L., ``A Plane Wave Expansion of Spherical Wave Functions for Modal Analysis
of Guided Wave Structures and Scatterers", \textit{IEEE Trans. on
Antennas and Propagation}, vol. 51, No. 10, pp. 2801-2805, 2003.

\bibitem{KThorne80}Thorne, K.S., ``Multipole expansions of gravitational radiation",
\textit{Reviews of Modern Physics}, vol.52, pp. 299-339, 1980.

\bibitem{AO96}Allen, B. and Ottewill, A.C., \textit{Detection of Anisotropies in the
Gravitational-Wave Stochastic Background}, University of
Wisconsin, and University College Dublin 1996.

\bibitem{Watson}Watson, G.N., \textit{A treatise on the theory of Bessel Functions}, Cambridge
University Press, 1958; Proc. Camb. Phil. Soc. vol. XIX, 96, 1918.

\bibitem{M1} Meissel, D.F.E,``Neue Entwickelungen ueber die Bessel'schen Functionen"\textit{Astr. Nach.} vol. CXXIX
col. 281-284, 1892; \textit{Astr. Nach.} vol. CXXX , ``Weitere
Entwickelungen ueber die Bessel'schen Functionen", col. 363-368,
1892; ``Einige Entwickelungen die Bessel'schen I-Functionen
betreffend" vol. CXXVII, col. 359-362, 1891; ``Beitrag zur theorie
der allgemeinen Bessel'schen Functionen" vol. CXXVIII (1891),
col.145-154.

\bibitem{PM} Peters P. C. and Mathews J., ``Gravitational Radiation from Point Masses in a Keplerian Orbit",
\textit{Phys. Rev.} vol. 131, pp. 435-439, 1963; Barack, L. and
Cutler, C., ``LISA Capture Sources: Approximate Waveforms,
Signal-to-Noise Ratios, and Parameter Estimation Accuracy"
\textit{Phys.Rev. D} vol. 69, 082005, 2004.

\bibitem{DBY}Debye, P., ``Naeherungsformeln fuer die Zylinderfunktionen fuer grosse Werte des
Arguments und unbeschraenkt veraenderliche Werte des Index",
\textit{Math. Ann.}, vol. LXVII pp.535-558, 1909.

\bibitem{AIR} Airey, J.R.,``Bessel and Neumann Functions of Equal Order and Argument" \textit{Phil. Mag.} (6) vol. XXXI, 520-528, 1916.


\end{thebibliography}
\end{document}